\documentstyle[preprint,aps]{revtex}

\begin{document}

\begin{center} The Original Mixed Symmetry States - $6^{+}_{1}$ 
and $6^{+}_{2}$ in $^{48}$Ti \medskip \\ 
Shadow J.Q. Robinson and Larry Zamick

\noindent  Department of Physics and Astronomy,\\
Rutgers University, Piscataway, \\New Jersey  08854-8019
\end{center}

\bigskip
\begin{abstract}
The $6^{+}_{1}$ and $6^{+}_{2}$ in $^{48}$Ti form a nearly degenerate
doublet.  In a single j shell calculation with the matrix elements
from experiment the $6^{+}_{1}$ changes sign under the interchange of
protons and neutron holes (odd signature) while the $6_{2}^{+}$ does
not (even signature).  As a consequence the calculated B(E2)
$6_{1}^{+}\rightarrow 4_{1}^{+}$ is much stronger than the
$6_{2}^{+}\rightarrow 4_{1}^{+}$ and the Gamow-Teller matrix element
to the $6_{2}^{+}$ state vanishes.  When using some popular
interaction e.g. FPD6 in single j shell the ordering of the even
signature and odd signature states gets reversed, so that the
Gamow-Teller matrix element to the $6^{+}_{1}$ state vanishes and the
$6_{2}^{+}\rightarrow 4_{1}^{+}$ E2 transition is the strong one.
When configuration mixing is introduced, the E2 transition
$6_{2}^{+}\rightarrow 4_{1}^{+}$ persists in being large. However the
Gamow-Teller strengths reverse, with the large matrix element to the
$6_{1}^{+}$ state in agreement with experiment.  Static properties
$\mu$ and Q for the two $6^{+}$ states are also considered.  The
experimental B(E2)'s from the $6^{+}$ states to the $4_{1}^{+}$ state
are not well known.
\end{abstract}
\vspace{2.5in}

\newpage

\section{Introduction}
In the early 1960's single j shell calculations in the f$_{7/2}$
region were performed by McCullen, Bayman, and Zamick (MBZ) [1,2] and
Ginnocchio and French[3].  In these calculations the two body matrix
elements were taken from experiment.  However the T=0 neutron proton
spectrum in $^{42}$Sc was not well determined.  Calculations with
correct T=0 matrix elements were later performed by Kutschera, Brown,
and Ogawa[4].

In single j shell calculations the energy levels and column vectors
are unchanged when protons and neutron holes are interchanged.[1,2,3]
This is known as cross conjugate symmetry.  For example in the
$f_{7/2}$ shell $^{43}$Sc (2 neutrons - one proton) and $^{53}$Fe (2
proton holes - neutron holes) are cross conjugate pairs.  It was noted
by MBZ [1,2] that the nucleus $^{48}$Ti is self-conjugate under this
transformation.  Thus if the wave function is written as
\begin{equation}
\psi = \Sigma
D(J_{P},J_{N})[(j^{2}_{\pi})^{J_{P}}(j^{-2}_{\nu})^{J_{N}}]^{J}
\end{equation}
for a given state either $D(J_{P},J_{N})=+ D(J_{N},J_{P})$ (even
signature) or $D(J_{P}J_{N})=- D(J_{N},J_{P})$ (odd signature).

The E2 matrix element from a state of even signature to that of odd
signature is proportional to $(e_{p} + e_{n})$ where $e_{p}$ and
$e_{n}$ are the effective E2 charges.  Between states of the same
signature the matrix element is proportional to $(e_{p}-e_{n})$. Thus
one expects large E2 transitions between states of opposite
signatures.  Note that static quadrupole moments in this simple model
are proportional to $(e_{p}-e_{n})$ and are thus not collective.

Another selection rule concerns the beta decay from the $6^{+}_{1}$
state of $^{48}$Sc (the isospin partner of a T=3 state in $^{48}$Ti) to
$6^{+}$ states in $^{48}$Ti - namely that the matrix element vanishes
to final states of even signature.

Since in this model there are no two body matrix elements between
states of even signature and odd signature there need not be any level
repulsion.  Indeed the $6^{+}_{1}$ and $6^{+}_{2}$ are very close in
energy both experimentally and in single j shell calculations.
However when more than one shell is involved, i.e. configuration
mixing, the whole concept of cross conjugate symmetry breaks down.
One can have mixing of what were dominately even signature and odd
signature states.  Do we still get a nearly degenerate $6^{+}$
doublet?  What happens to the selection rules? We will consider these
questions in the next section.

\section{Calculation}
We present detailed results for the properties of the $6^{+}_{1}$ and
$6^{+}_{2}$ states in $^{48}$Ti in Table I.  We give the energies, the
energy splittings, the magnetic moments, the electric quadrupole
moments, the B(E2) rates from the $6^{+}$ states to the $4_{1}^{+}$
state, and the values of B(GT) from the beta decay of the $6^{+}$
ground state of $^{48}$Sc to the two $6^{+}$ states in $^{48}$Ti.  For
the electic quadrupole properties we use effective charges of 1.5 for
the protons and 0.5 for the neutrons.  We do not use renormalized
Gamow-Teller operators although an argument can be made that the
transitions rates should be multiplied by about 0.5.

We first present results of a single j shell calculation (j=f$_{7/2}$)
in which the two body matrix elements are taken from the spectrum of
$^{42}$Sc, for which the excitation energies of the J=1 to J=7 states
are respectively 0.6110 MeV, 1.5863 MeV, 1.4904 MeV, 2.8153 MeV,
1.5101 MeV, 3.242 MeV, and 0.6163 MeV.

We then perform calculations with the FPD6 interaction[5] in which up
to t particles are excited from f$_{7/2}$ to higher shells.  We first
give results for t=0.  This is the same space as the one in which the
spectrum of $^{42}$Sc is used as input.  The results however are
\underline{qualitatively} different because the interaction is
different.  We then study the effects of configuration mixing at the
t=1 and t=2 levels for the electromagnetic and weak properties and we
also perform a t=3 calculation for the energy splitting of the two
$6^{+}$ states.

Let us first compare the two t=0 calculations.  We see that there is a
qualitative difference.  In the calculation in which matrix elements
are taken from experiment, the $6_{1}^{+}$ state has the odd
signature.  This is evident from the fact that the Gamow-Teller matrix
element to this state is not zero. and the B(E2) from this state to
the $4_{1}^{+}$ state which has even signature is strong.  The
$6_{2}^{+}$ state in this calculation has even signature - the value
of B(GT) to this state from the $^{48}$Sc($6_{1}^{+}$) beta decay
\underline{is} zero and the B(E2) to the $4_{1}^{+}$ state is very
weak.

Just to clarify the E2 rules, note that the J=$0_{1}^{+}$ state
necessarily has even signature.  The sequence of B(E2)'s is then
$0_{1}$ (even) $\rightarrow$ $2_{1}$ (odd) $\rightarrow$ $4_{1}$
(even) $\rightarrow$ $6_{1}$ (odd).  Thus when the matrix elements
are taken from the spectrum of $^{42}$Sc in a single j shell
calculation, the strong B(E2) sequence is $0_{1} \rightarrow 2_{1}
\rightarrow 4_{1} \rightarrow 6_{1}$.

When the FPD6 matrix elements are used in a t=0 calculation B(GT) to
the second $6^{+}$ state vanishes, and B(E2) from the second $6^{+}$
state to the $4_{1}^{+}$ is the strong one.  Thus it is clear that
with FPD6 t=0 the order of the $6^{+}$ states is the reverse of that
when the spectrum of $^{42}$Sc is used as input.  For FPD6, the
$6^{+}_{1}$ state has even signature and the $6_{2}^{+}$ state odd
signature.

When configuration mixing is introduced the states no longer have
definite signature.  What now?  We first look at the
$6_{2}$-$6_{1}$ energy splitting.  We now have level repulsion so it
is not clear if the energy splitting will continue to be small.

For t=0,1,2, and 3 using the FPD6 interaction [5] the calculated
splittings are respectively 0.076, 0.156, 0.229 and 0.161 MeV.  The splitting
continues to be small despite the level repulsion and indeed the t=3
result agrees well with the experimental value of 0.175 MeV.[6]

Let us now consider B(E2) and B(GT).  As we go from t=0 to t=2 there
is no change in the \underline{qualitative} fact that B(E2) $6_{2}
\rightarrow 4_{1}$ is much stronger than B(E2) $6_{1} \rightarrow
4_{1}$.  The t=0 results for $6_{1} \rightarrow 4_{1}$ $6_{2}
\rightarrow 4_{1}$ are 0.935 and 43.54 $e^2fm^{4}$, while for t=1 they
are 26.63 and 53.07 $e^2fm^{4}$ and for t=2 they are 16.13 and 65.56
$e^2fm^{4}$.  We emphasize that these results are in
\underline{qualitative disagreement} with the single j shell
calculation using matrix elements taken from experiment.  For that
calculation the B(E2) from the $6_{1}^{+}$ state to the $4_{1}^{+}$ is
the strongest.  We will discuss the expermential situation later.

Let us now look at B(GT) from the $6^{+}_{1}$ state of $^{48}$Sc to
the $6^{+}$ states in $^{48}$Ti.  Here the evaluation with increasing
t is different than in was for B(E2).  For FPD6 at the t=0 level the
B(GT) to the $6^{+}_{1}$ state is zero because in this calculation the
$6_{1}^{+}$ state has even signature.  However at the t=2 level, we
find a reversal with B(GT) to the $6_{1}^{+}$ state larger than to the
$6_{2}^{+}$ state.  The respective values are 0.09559 and 0.01104.

The fact that the value of B(GT) is larger to the $6_{1}^{+}$ state is
in agreement with experiment.[6]  The experimental values are log ft =
5.563 to the $6_{1}^{+}$ state and log ft = 6.010 to the $6_{2}^{+}$
state.  The difference is in log ft is 0.447.  The calculated
difference from Table I (t=2) is 0.938.  But the results seem to be
changing rapidly with t.

Getting back to the B(E2)'s, unfortunately they are poorly known.[4]  For
the decay $6_1 \rightarrow 4_1$ the uncertainty is very large. $0.221
ps \leq T_{1/2} \leq 8.9 ps$ while for $6_2 \rightarrow 4_1$ we have
$1.4 ps\leq T_{1/2} \leq 2.4 ps.$ Clearly it would be of great interest to
have better measurement.

\section{A quasi-rotational picture- Static properties}

What appears to be emerging as we increase the configuration space is
that we are approaching but not fully reaching a situation where a
rotational description can be used.  In particular it is useful to
consider the K quantum numbers.  It would appear that the $6_{2}^{+}$
state can best be described as belonging to a ground state band where
the dominant quantum number is K = 0.  The other members would be
$0_1^+$, $2_1^+$ and $4_1^+$ states The $6_{1}^{+}$ would then have a
higher K value.

This is supported by the beta decay data.  In $^{48}$Sc we have an odd
proton in a K=1/2 state while the odd neutron would be in K=7/2.  From
this we could form K=3 or 4.  For allowed Gamow-Teller transitions
$\Delta K = \pm $1 or 0 so a transition to a K=0 state is forbidden.
Amusingly then what starts out as a single j shell selection rule
(signature) changes in a K selection rule.

Let us consider the static quadrupole moments.  As we increase t from
0 to 2 the value of $Q(6_{2}^{+})$ changes from -0.892 to - 23.31 to
-26.15 $efm^2$.  The large negative Q is consistent with a prolate
intrinsic quadrupole moment
\begin{equation}
Q=\frac{(3 K^2-J(J+1))Q_{0}}{(2J+3)(J+1)}
\end{equation}
For K=0 J=6 we get Q=-6/13 $Q_{0}$.

The values for Q for the $6_1^+$ state for t=0,1, and 2 are
respectively -2.176 to 21.34 to 25.42 $e fm^2$.  The final value is
almost equal but opposite to the quadrupole moment for the $6_{2}^{+}$
state.  It probably corresponds to an oblate state but the K quantum
number is difficult to ascertain.

However the magnetic moment of the $6_{2}^{+}$ state does not fit a
pure K=0 picture.  In that picture $\mu$=6 Z/A =2.75nm but the
calculated values for t=0,1, and 2 respectively are 3.326, 7.392, and
7.451 nm,.  Hence even if the $6_{2}^{+}$ state is dominately a K=0
state there must be higher K admixtures.

Note that for the $6_1^+$ state $\mu$ changes from 3.326 to -1.332 to
-1.914 nm as we increase t from 0 to 2.

In closing we repeat the fact that there are \underline{qualitative}
as well as quantitative differences between single j shell
calculations in which the matrix elements are taken from experiment
and those in which configuration mixing is included.  In the former it
is the $6^{+}$ state which has the largest B(E2) to the $4_{1}^{+}$
state which has the largest GT matrix element from the J=6$^{+}_{1}$
in $^{48}$Sc.  In the latter it is the state with the smallest B(E2)
to the $4_{1}^{+}$ state which has the largest GT matrix element.  In
the former calculation it is the 6$_{1}^{+}$ state which has the
strongest B(E2) while in configuration mixing it is 6$_{2}^{+}$.

However there is a point of agreement.  In both calculations the near
degeneracy of the 6$^{+}$ states in maintained.  This is somewhat of a
surprise because in configuration mixing, there should be level
repulsion.

This work was supported by the U.S. Dept. of Energy under 
Grant No. DE-FG02-95ER-40940 and one of us by 
a GK-12 NSF9979491 Fellowship(SJQR).

\begin{table}
\caption{Level and Electro-weak data}
\begin{tabular}{ccccccc}
\tableline
Calculation$^{a}$  & J$^{\pi}$ & E(MeV)& $\mu$ (n.m) & Q (e fm$^{2}$) 
& B(E2 $\rightarrow $ 4$_{1}^{+}$)e$^{2}$fm$^{4}$ & B(GT $\rightarrow$ 6) \\
\tableline
Experimental    & 6$_{1}^{+}$ & 0.000  & 3.329 & -0.462 & 44.82 & .09175      \\
Matrix Elements & 6$_{2}^{+}$ & 0.016  & 3.322 & -1.719 & 0.841 &  0           \\
from $^{42}$Sc t=0  & 4$_{1}^{+}$ & -0.761       & 2.217 &  4.567 &       &            \\
FPD6 t=0 & 6$_{1}^{+}$ & 0.000  & 3.326  & -2.176 & 0.935 & 0                 \\
         & 6$_{2}^{+}$ & 0.076  & 3.326  & -0.892 & 43.54 & 0.1631            \\
         & 4$_{1}^{+}$ & -0.457       & 2.217  & 3.920  &       &                   \\
FPD6 t=1 & 6$_{1}^{+}$ & 0.000  & -1.332 & 21.34  & 26.63 & 0.09953           \\
         & 6$_{2}^{+}$ & 0.156  &  7.392 & -23.31 & 53.07 & 2.869 10$^{-5}$ \\
         & 4$_{1}^{+}$ &-0.529        &  2.258 & -6.126 &       &                   \\
FPD6 t=2 & 6$_{1}^{+}$ & 0.000  & -1.914 &  25.42 & 16.13 & 0.09569           \\
         & 6$_{2}^{+}$ & 0.229  &  7.451 & -26.15 & 65.56 & 0.01104           \\
         & 4$_{1}^{+}$ &-0.584        &  1.722 & -10.51 &       &                   \\         
\end{tabular}
a For t=3 the 6$_{2}^{+}$-6$_{1}^{+}$ splitting decreases to 0.161 MeV
\end{table} 

\centerline{References}
\bigskip

\begin{enumerate}
\item B.F. Bayman, J.D. McCullen and L. Zamick, Phys Rev Lett
\underline{10}, (1963) 117
 
\item J.D. McCullen, B.F. Bayman, and L. Zamick, Phys Rev
\underline{134}, (1964) 515; ``Wave Functions in the 1f$_{7/2}$
Shell'', Technical Report NYO-9891

\item J.N. Ginocchio and J.B. French Phys. Lett \underline{7}, (1963)
137

\item W. Kutschera, B.A. Brown and K. Ogawa Riv. Nuovo Cimento
\underline{1}, (1978) 12

\item W.A. Richter, M.G. Van der Merwe, R.E. Julies and B.A. Brown,
Nucl. Phys. \underline{A253}, (1991) 325

\item T.W. Burrows, Nuclear Data Sheets \underline{68}, (1993) 1

\end{enumerate}

\end{document}